\documentclass[twocolumn]{aastex62}

\usepackage{latexsym}
\usepackage{color}
\usepackage{amsmath}
\usepackage{amssymb}
\usepackage{graphicx}
\usepackage{aas_macros}

\newcommand{\Ms}{M$_{\odot}$}

\newcommand{\Lya}{Lyman\,$\alpha{}$}

\graphicspath{{./}{figures/}}

\submitjournal{ApJL}

\shorttitle{Pop III Constraints from 21cm-Cosmology}
\shortauthors{Schauer et al.}

\begin{document}

\title{Constraining First Star Formation with 21cm-Cosmology}

\correspondingauthor{Anna T. P. Schauer}
\email{anna.schauer@utexas.edu}

\author[0000-0002-2220-8086]{Anna T. P. Schauer}
\altaffiliation{Hubble Fellow}
\affiliation{Department of Astronomy, 
University of Texas at Austin,
TX 78712, USA}

\author{Boyuan Liu}
\affiliation{Department of Astronomy, 
University of Texas at Austin,
TX 78712, USA} 

\author{Volker Bromm}
\affiliation{Department of Astronomy,
University of Texas at Austin,
TX 78712, USA} 

\begin{abstract}
Within standard $\Lambda$CDM cosmology, Population~III (Pop~III) star formation in minihalos of mass $M_\mathrm{halo}\gtrsim 5\times10^5$\Ms\,provides the first stellar sources of \Lya\,(Ly$\alpha$) photons. The Experiment to Detect the Global Epoch of Reionization Signature (EDGES) has measured a strong absorption signal of the redshifted 21\,cm radiation from neutral hydrogen at $z\approx 17$, requiring efficient formation of massive stars before then. In this paper, we investigate whether star formation in minihalos plays a significant role in establishing the early Ly$\alpha$ background required to produce the EDGES absorption feature. 
We find that Pop~III stars are important in providing the necessary Ly$\alpha$-flux at high redshifts, and derive a best-fitting average Pop~III stellar mass of 
$\sim$\,750\,\Ms{}\,per 
minihalo, corresponding to a star formation efficiency of 0.1\%. Further, it is important to include baryon-dark matter streaming velocities 
in the calculation, to limit the efficiency of Pop~III star formation in minihalos. Without this effect, the cosmic dawn coupling between 21\,cm spin temperature and that of the gas would occur at redshifts higher than what is implied by EDGES. 
\end{abstract}


\section{Introduction}
The recent detection of a 21\,cm signal at high redshift has opened a new window for astrophysics at the dawn of star formation \citep{furlanetto06,pl12}. EDGES has measured a strong, global (sky-averaged)
absorption signal centered around 78\,MHz \citep{bow18}. 
The absorption signal is broad, and a factor of about three stronger
than expected within standard $\Lambda$CDM, where dark matter only interacts gravitationally. If verified,  that signal points to new dark matter physics \citep[e.g.,][]{bar18,ml18,slatyer18,fialkov18}, or an additional radio background \citep{ew18,fh18}.

In this study, we focus on a second characteristic, the implied timing of early star formation. 
The absorption signal starts at $z\approx 20{}$ and is strongest at
$z \approx 17$, indicating that at that time, the spin temperature
of neutral hydrogen is tightly coupled to the gas temperature.

This coupling is mediated through Ly$\alpha$-radiation via the Wouthuysen-Field effect \citep{w52, field58}. The critical Ly$\alpha$ background intensity required for effective coupling has been estimated to be $1.8\times 10^{-21}\,
[(1+z)/20]$\,erg\,s$^{-1}$\,cm$^{-2}$\,Hz$^{-1}$\,sr$^{-1}$
\citep{mmr97,cm03}. 
Pop~III stars are typically more massive and therefore hotter than standard populations, resulting in an increased Ly$\alpha$\,luminosity \citep{volkerreview13,glov13}.
The role of X-ray sources in shaping the thermal history of the early intergalactic medium (IGM) is still uncertain \citep[e.g.,][]{jeon14,jeon15}, and we thus neglect their contribution in this study.

The Ly$\alpha$\,flux emitted from the first galaxies has been studied before in the context of EDGES,  requiring large star formation efficiencies to allow strong coupling before redshift $z\simeq17$ \citep[e.g.,][]{madau18,mf18}. Here, we test
whether Pop~III stars in minihalos significantly contribute to the overall Ly$\alpha$\ luminosity, and whether the combined star formation activity at high $z$ can provide the necessary photon flux at the right time. Our analysis also provides an upper limit on the overall Pop~III star formation efficiency (SFE), as star formation cannot occur too early.

We further include a crucial large-scale effect that influences Pop~III star formation in minihalos, the relative motion between the cosmic baryon and dark matter components. These streaming velocities date back to the epoch of recombination \citep{th10}, described by a multi-variate Gaussian spatial distribution with a standard deviation of $\sigma_\mathrm{rms} = 30$\,km\,s$^{-1}$. The initially supersonic motion can be assumed to be coherent over large (Mpc) scales, decaying as the Universe is expanding. One key effect is the reduced baryon fraction in halos located within regions with streaming velocities, which subsequently leads to a reduced halo mass function \citep[e.g.,][]{naoz12,fialkovreview14}. As a result, star
formation in such regions is delayed
\citep[e.g.,][]{greif11,stacy11a,naoz13,hirano18,schauer18}, and the halo
mass necessary for Pop~III star formation increases. When estimating the star formation rate and Ly$\alpha$-background flux, we include these effects in our modelling.
%
\section{Methodology}
\label{sec:outline}
From recent simulations \citep{schauer18}, we know the average minihalo mass $M_\mathrm{ave}$ necessary for star formation, depending on the halo's streaming environment. Using the Sheth-Torman mass function \citep{st01}, we then estimate the respective number of halos that have crossed the mass threshold for star formation. With streaming motions distributed according to a three dimensional Gaussian
\citep{th10}, we can calculate the fraction of the Universe exposed to a given streaming velocity. Convolving these results, we arrive at an estimate for the number density of star forming halos, as a function of mass and redshift.

In a second step, we parametrize the star formation efficiency for these sources, distinguishing between a Pop~III and Pop~II stellar component, as the first, metal-free stars are more massive and therefore more luminous \citep[e.g.,][]{volkerreview13,glov13}. Since star formation in
minihalos is very bursty, we assume a one-time star formation event 
in each Pop~III host minihalo, 
and a fixed star formation history in the more massive halos that host Pop~II stars. Finally, we calculate the Ly$\alpha$\ background luminosity, based on the global stellar density. We provide best-fitting, combined Pop~III/Pop~II models that match the redshift position of the EDGES signal.

\section{Results}
\label{sec:results}
\subsection{Threshold Masses for Pop~III Hosts}
We base our analysis on minihalos formed in the \cite{schauer18} high-resolution cosmological simulations \citep[see also][]{anna17b}. 
The simulations are initialized at redshift $z=200$ with {\it Planck} 
parameters \citep{Planck15}, performed with the AREPO code \citep{arepo}, including a network of primordial chemistry. 
To represent different streaming regions, a constant offset velocity is added to the initial conditions, with an amplitude of 0, 1, 2 and 3 $\sigma_\mathrm{rms}$ (v0, v1, v2, and v3). For the 3 $\sigma_\mathrm{rms}$ case, a bigger box with four times longer side length is also run (v3\_big). 

\begin{figure*}
\centering
\includegraphics[width=2.\columnwidth]{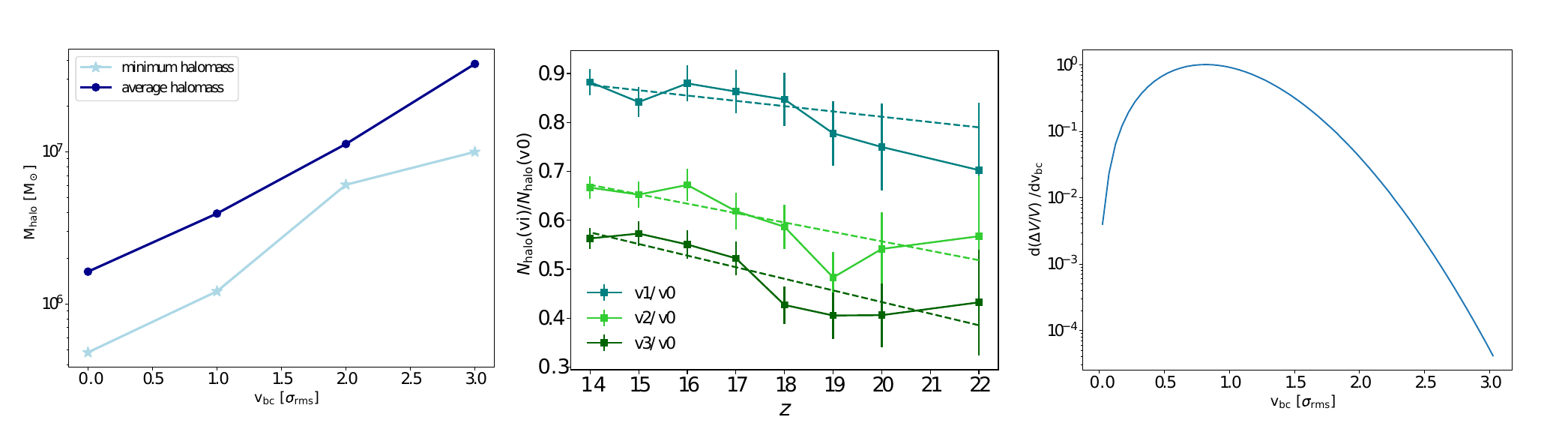}
\caption{Effects of streaming motion.
{\it Left panel:} Minimum (light blue stars) and average (dark blue dots) halo mass for star formation in minihalos, as a function of streaming
velocity (independent of redshift), based on \protect{\cite{schauer18}}, for their
v0, v1, v2, v3\_big simulations.
{\it Middle panel:} Relative number of halos for different streaming velocities, as a function of $z$, considering all halos with masses larger than $M_\mathrm{halo}\ge 4.8\times10^5$\,\Ms. The solid lines show the raw data, and the dashed lines our redshift dependent linear fits.  
{\it Right panel:} Volume fraction as a function of streaming velocity, 
in units of $\sigma_\mathrm{rms}$ (independent of redshift).}
\label{fig:mvbc}
\end{figure*}

In Fig.~\ref{fig:mvbc}, we illustrate the different effects of streaming velocities. Specifically, in the left panel we show the minimum and average halo masses for the corresponding gas to become cold and dense in the center, and hence eligible for star formation, as a function of streaming velocity. \cite{schauer18} do not see any evolution in the minimum or average halo mass, and we employ their redshift-independent threshold values. In the following, we work with the
average halo mass, above which more than 50\% of all halos are star forming. 

%
\subsection{Halo Number Densities}
\label{sec:stream_con}
We utilize the halo mass function python tool \textit{hmf} \citep{hmf} to
derive the halo number density for star forming minihalos, $N(M \ge M_\mathrm{ave}){}$, as a function of redshift and halo mass, with the same \cite{Planck15} cosmological parameters as \cite{schauer18}. We chose the Sheth-Torman mass function\footnote{We have verified that our main conclusions do not change for a Press-Schechter function}, which is generally thought to fit better at high redshifts \citep{reed07}\footnote{We chose to work in this semi-analytical framework, as fitting functions from simulations depend on the cosmology and are rare at high redshifts \citep[see][for a WMAP cosmology]{trac15}.}. 
As evident in the middle panel of Fig.~\ref{fig:mvbc}, the halo mass function is reduced in regions with streaming velocity. Specifically, we show the fraction of the mass functions in streaming regions (v$i$), relative to the no-streaming (v0) case, considering all
halos with $M_\mathrm{halo}\ge 4.8\times10^5$\,\Ms. This mass limit corresponds to the minimum halo mass for star formation in the v0 simulation. The halo number is only slightly reduced in simulation v1,
but significantly so, at the 50\%-level, in simulations v2 and v3. At higher redshifts, we have less data to sample as fewer halos have exceeded the mass threshold, resulting in larger error bars.

To estimate how common the streaming regions are, we derive their respective volume filling fractions \citep[e.g.,][]{tbh11,greif11,fialkovreview14}. The velocity distribution follows a multivariate Gaussian:
\begin{equation} P_\mathrm{vbc} (v_\mathrm{bc}) = \left(
\frac{3}{2\pi\sigma_\mathrm{rms}^2}\right)^{3/2} 4\pi v_\mathrm{bc}^2
\exp{\left( - \frac{3 v_\mathrm{bc}^2}{2 \sigma_\mathrm{vbc}}\right)} .
\end{equation} 
Integrating $P_\mathrm{vbc}{}$ to infinity, one can derive the fraction of 
the volume with $v_\mathrm{bc}{}$ or
higher. E.g., regions with streaming velocities of
2.0\,$\sigma_\mathrm{rms}$ or higher make up less than one percent of the cosmic volume. In the right panel of Fig.~\ref{fig:mvbc}, we present the differential volume fraction, which peaks around 0.8\,$\sigma_\mathrm{rms}$. We note that regions with very small streaming velocity are not common.

\begin{figure*}
\centering
\includegraphics[width=1.99\columnwidth]{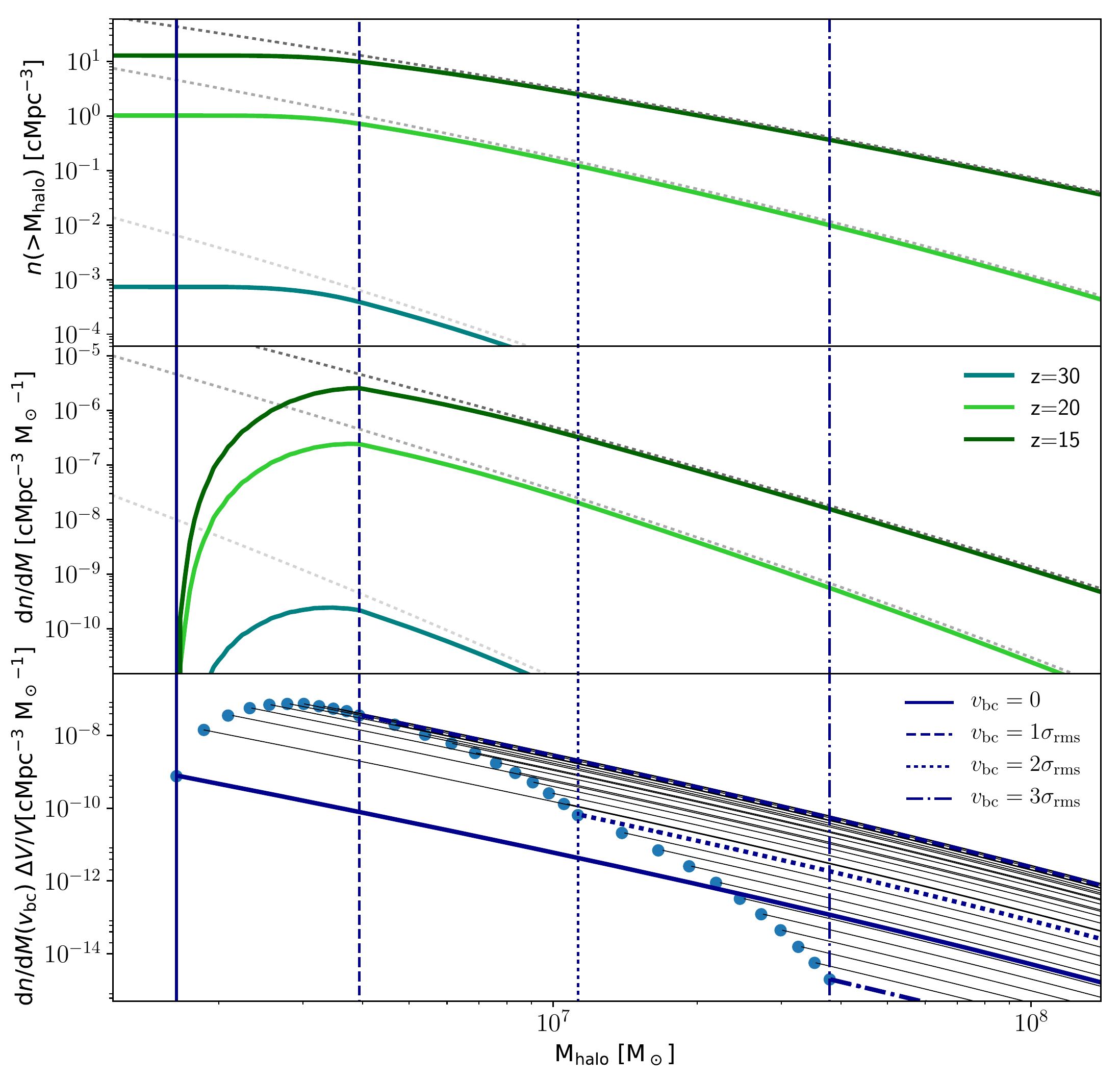}
\caption{Halo mass function. {\it Upper panel:} Volume-averaged cumulative mass functions for star forming minihalos for $z=30$ (turquoise), $z=20$ (green)
and $z=15$ (dark green), weighted by the volume fraction of different
streaming velocity regions. {\it Middle panel:} Differential mass function  for star forming minihalos for the same redshifts. In the upper two panels, the grey dotted lines show the situation without streaming velocities.
{\it Lower panel:} Mass functions for star forming minihalos for small intervals around $v_\mathrm{bc}$, with $\Delta\,v_\mathrm{bc} = \,0.1 \sigma_\mathrm{rms}$, evaluated at $z=20$. Select velocities are highlighted with the dark blue lines. In all three panels, the vertical lines show the halo mass limit for star formation for the v0 (solid line), v1 (dashed line), v2 (dotted line) and v3 (dash-dotted line) cases.  }
\label{fig:hmf_vbc}
\end{figure*}

We present the results for the mass function of star forming minihalos in
Fig.~\ref{fig:hmf_vbc}, with values given in comoving units. In the lower panel, we show the differential halo mass function for different streaming regions at redshift $z=20$. For a given streaming velocity, we apply the corresponding minimum mass cut (shown by the blue dots), and multiply with the respective volume filling fraction for an interval 
$v_\mathrm{bc}- \Delta v_\mathrm{bc}$ to $v_\mathrm{bc} + \Delta v_\mathrm{bc}$. A volume filling fraction of 99\% corresponds to a mass threshold of $2.1\times10^6$\,\Ms. The black lines represent the resulting differential mass
functions in that streaming region, and their sum equals the volume averaged halo mass function, as shown in the middle panel. One can see that the contribution to the mass function is largest around $v_\mathrm{bc} \approx 1 \sigma_\mathrm{rms}{}$. Combining these results, we derive the cumulative halo mass function for star forming minihalos, averaged over the various streaming regions, shown in the upper panel of Fig.~\ref{fig:hmf_vbc}
for redshifts $z=15$, 20 and 30. When accounting for streaming velocities, we find up to one order of magnitude fewer Pop~III star forming minihalos, with a comparable effect on the high-$z$ Ly$\alpha$ background flux. It is therefore important to include the impact of streaming motions in any realistic modelling.

\subsection{Star Formation Models}
\label{sec:lya}
Star formation at high redshift is typically very bursty. After the first stars have formed in a minihalo, their feedback can prevent further star formation until the halo has grown to higher masses and new gas has fallen in \citep[e.g.,][]{pmb12}. For higher mass halos, however, continuous star formation is possible, from gas that is already metal-enriched.
\begin{figure}
\centering
\includegraphics[width=1.1\columnwidth]{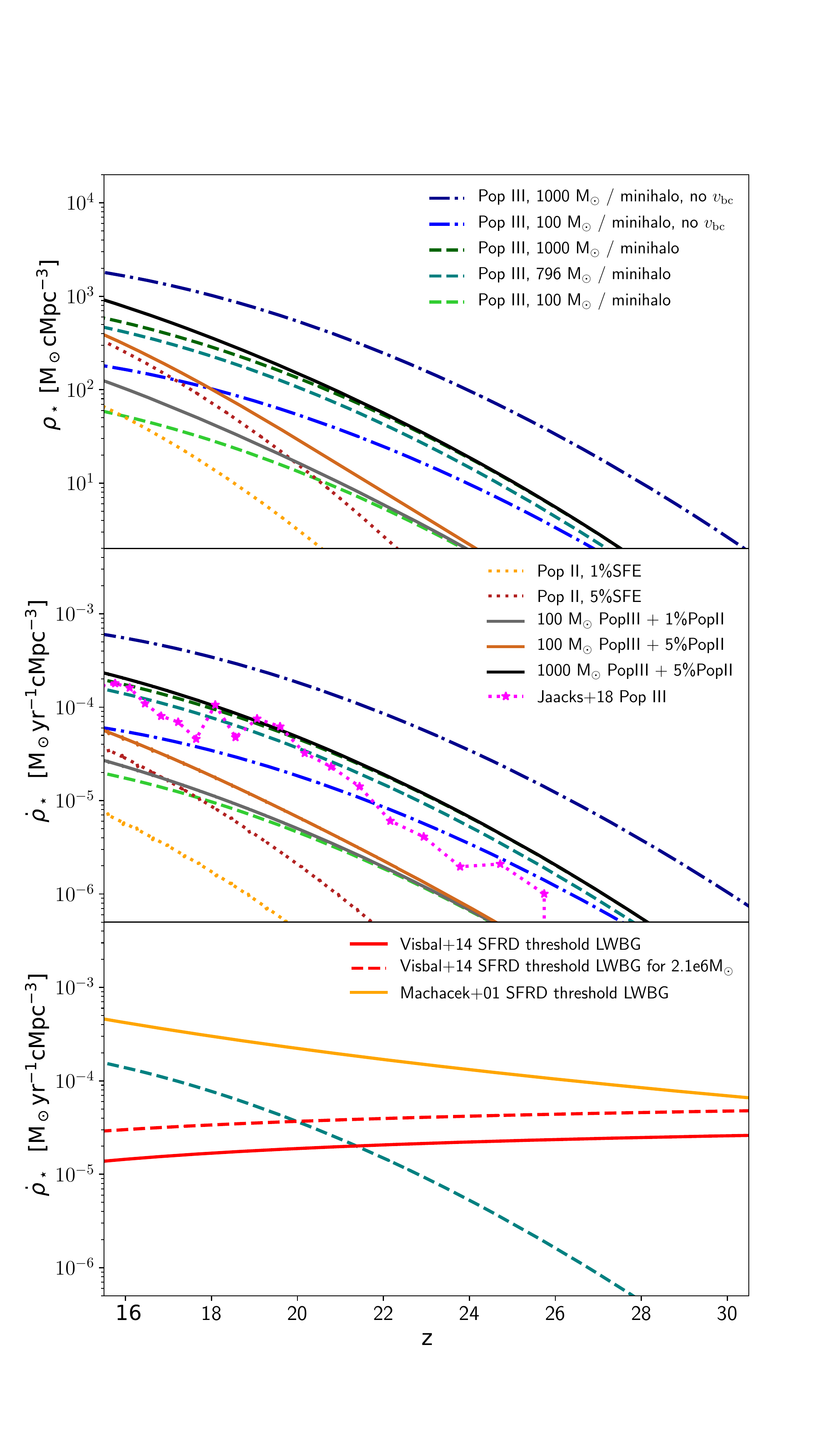}
\caption{{\it Upper panel:} Comoving density in massive stars as a function of redshift.
{\it Middle panel:} Comoving star formation rate density. 
We show the results for a Pop~III component (dashed lines) and
a Pop~II component (dotted lines) alone, as well as the combined models
(solid lines). The dot-dashed lines show Pop~III models when
neglecting streaming velocities. The magenta line shows the star formation rate density from \cite{jaacks18} for comparison. {\it Lower panel:} Comoving star formation rate density above which a LWBG provides the stronger constraints on the minimum halo mass. }
\label{fig:mdot}
\end{figure}
To mimic this dependence on halo mass, we apply a mass threshold, where bursty Pop~III star formation transitions to a near-continuous mode.
We here make the implicit assumption of instantaneous star formation once the mass threshold is crossed. \cite{pmb12} have found in their simulations that the transition in star formation mode occurs at $M_\mathrm{halo}^\mathrm{thres} \approx
1\,\times\,10^8$\,\Ms, and we adopt this value, for simplicity assumed to be redshift-independent. We summarize our model in Table~\ref{tab:models}, where star formation efficiencies are free
parameters. In addition, we consider a comparison model without streaming velocities.
\begin{table*}
\begin{center}
\begin{tabular}{l|cccc}
Model   & $M_\mathrm{halo}$                                     & SFE                                    & $\dot N_\mathrm{ion}$           & $t_\star$ \\
\hline
Pop~III & $M_\mathrm{ave} < M_\mathrm{halo} < 1\times10^8$\,\Ms & $M_\star^\mathrm{III}$ per halo              & 10$^{48}$\,s$^{-1}$\,\Ms$^{-1}$ & 3\,Myr \\
Pop~II  & $M_\mathrm{halo} > 1\times10^8$\,\Ms                  & SFE = $M_\star^\mathrm{II} / M_\mathrm{gas}$ & 10$^{47}$\,s$^{-1}$\,\Ms$^{-1}$ & 10\,Myr \\
\hline
\end{tabular}
\caption{Star formation parameters. $M_\star^\mathrm{III/II}$ is the total stellar mass in Pop~III/II, and $t_\star$ the typical lifetime of massive stars.}
\label{tab:models}
\end{center}
\end{table*}
We calculate the total physical star formation rate density, $\dot\rho_\star$, as a
combination of the Pop~III and Pop~II components:
\begin{eqnarray}
\dot\rho_\star^\mathrm{III} &=& M_\star^\mathrm{III} \,\frac{\mathrm{d}N_\mathrm{halo}}{\mathrm{d}t \,\mathrm{d}V} \\
\dot\rho_\star^\mathrm{II}  &=& \mathrm{SFE}\, f_b \, \frac{\mathrm{d}M_\mathrm{halo}}{\mathrm{d}t \, \mathrm{d}V}, 
\end{eqnarray}
where $f_b = \Omega_b / \Omega_0 = 0.16{}$ is the halo gas fraction, assumed to be equal to the global baryon fraction.

Our results are presented in Fig.~\ref{fig:mdot}, with (comoving) star formation rate densities shown in the middle panel. We choose values of 100\,\Ms, 1000\,\Ms, as well as our later determined best-fit model with $\sim$\,800\,\Ms\,in Pop~III stars formed in a minihalo, and Pop~II SFEs of 1\% and 5\%. One can see that
Pop~III stars initially dominate, with Pop~II star formation becoming important after $z\approx 20$. Neglecting streaming velocities
(dot-dashed lines) leads to unphysically high values. 

In the lower panel, we assess at which star formation rate densities the mass threshold dependent on a LWBG becomes more strict than our halo mass threshold. Hereby, we convert the star formation rate density to a LW flux based on a calculation by \cite{johnson11}. Then we derive the halo mass threshold for star formation based on the LWBG flux with two literature models from \cite{met01} and \cite{vis14}. In this way, for any given SFRD, we can calculate the mass threshold set by the corresponding LWBG, and evaluate whether the effect of a LWBG is dominant by comparing the LWBG-based mass thresholds with those dictated by streaming velocities. We chose the mass thresholds of 1.6$\times10^6$\,\Ms and 2.1$\times10^6$\,\Ms to compare with, corresponding to our minimum halo mass for 100\% and 99\% of the volume of the Universe (based on the streaming velocity distribution). One can see that our fiducial model lies below the \cite{met01} SFRD threshold at all times, and that the LWBG constraint from \cite{vis14} only takes over at redhshifts $z < 20{}$ for a volume fraction of 99\%. When considering streaming velocities, an additional LWBG does not change the SFRD at these high redshifts.

The corresponding total (comoving) mass densities in massive stars  can be obtained by integrating over the stellar lifetime (see Table~\ref{tab:models}), shown in the upper panel of Fig.~\ref{fig:mdot}. As massive Pop~II stars have $\sim 3$ times longer lifetimes, their contribution to the total stellar density is a factor of $\sim 3$ higher, compared to the star formation rate density.

\subsection{Ly$\alpha$ Background Flux}
\label{sec:lya}
The Ly$\alpha$ background intensity can be calculated by integrating over the photon sources in a cosmological volume large enough to allow photons to redshift into the Ly$\alpha$ line. For simplicity, we include all photons between the Ly$\alpha$ resonance and the hydrogen ionization
limit, resulting in $(1+z_\mathrm{max})/(1+z) = 4/3$. We further assume a pure blackbody and use effective temperatures corresponding to the ionizing photon numbers for Pop~III and Pop~II stars (see Table \ref{tab:models}).
Following \cite{cm03}, the Ly$\alpha$ background intensity from Pop~III stars is then
\begin{equation}
J_\alpha(z) 
= \frac{c}{4\pi} \int_z^{z_\mathrm{max} }
\mathrm{d}z^\prime \frac{\mathrm{d}t}{\mathrm{d}z^\prime}
\frac{(1+z)^3}{(1+z^\prime)^3}
j_\nu(z^\prime) . 
\end{equation}

The proper specific intensity $j_\nu(z^\prime)$ can be derived from the spectral energy distribution and the proper stellar mass density:
\begin{equation}
j_\nu(z^\prime) = L_\nu(z^\prime) \rho_\star(z^\prime)
= \frac{L}{\sigma_\mathrm{B} T^4} \pi B_\nu(T, z^\prime)\rho_\star(z^\prime), 
\end{equation}
where the luminosity can be approximated by the Eddington limit $L \approx L_\mathrm{Edd} = 1.25\times 10^{38} \mathrm{erg\,s}^{-1}(M/\mathrm{M}_\odot)$ \citep{bkl01}, $\sigma_\mathrm{B}$ is the Stefan-Boltzmann constant, $T$ the effective temperature of the black body, and $B_\nu(T, z^\prime)$ is the Planck function. $\rho_\star(z^\prime)$ is the proper stellar density of all Pop~III stars which have not exceeded their lifetime.
The calculation for $J_\alpha^\mathrm{II}{}$ follows
analogously.

\begin{figure}
\centering
\includegraphics[width=1.1\columnwidth]{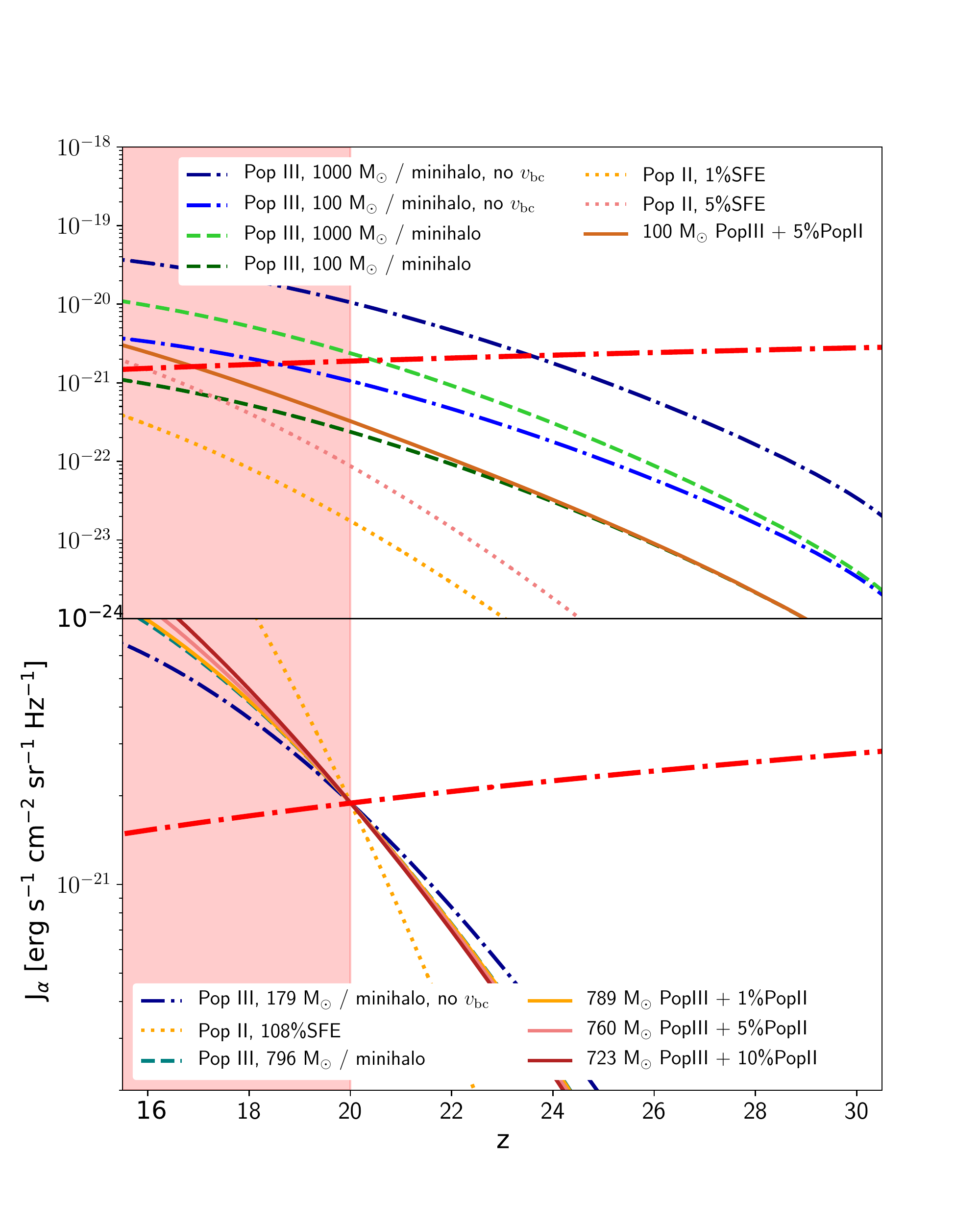}
\caption{Ly$\alpha$-flux as a function of redshift for
our different models.
Line styles and colors in the upper panel are the same as in Fig.~\ref{fig:mdot}.
In the lower panel, we show our best-fit models for a Pop~II component alone
(dotted orange line), a Pop~III component alone (dashed green line), a Pop~III
component that neglects streaming velocities (blue dot-dashed line), and best fit
models with a Pop~II SFE of 1, 5, and 10\% in combination with a Pop~III component (solid lines).
The red dot-dashed line shows the Ly$\alpha$-flux necessary for effective coupling
(from \protect{\citealt{cm03}}). The shaded red area on the left represents the EDGES timing constraint.}
\label{fig:nlya}
\end{figure}

The EDGES result implies that the spin temperature of neutral hydrogen needs to be efficiently coupled to the kinetic gas temperature for $z\lesssim 20$. This can be achieved when the thermalization rate due to Ly$\alpha$ scattering 
is stronger than the coupling between the spin temperature and the cosmic microwave background. Evaluating this condition, 
\cite{cm03} find that $J_\alpha \gtrsim 9 \times 10^{-23}(1+z)$\,erg\,s$^{-1}$\,cm$^{-2}\,$Hz$^{-1}$\,sr$^{-1}$ is required, and we use their estimate in our analysis (see Fig.~\ref{fig:nlya}).

In the lower panel of Fig.~\ref{fig:nlya}, we show the best fit models. For a Pop~II component alone, we would require an unphysical SFE of $> 100$\%. 
We thus conclude that Pop~III star formation is crucial in this context and should not be neglected. Alternatively, \cite{mf18}, who 
do not include minihalos, have to assume a steepening of the UV luminosity function at high redshifts. It is also important to include a treatment of streaming velocities, as one would otherwise predict efficient Wouthuysen-Field coupling too early in cosmic history, incompatible with the EDGES timing constraint. If we consider a combined Pop~III/II model with a plausible Pop~II SFE (less than 10\%), we infer a best-fit average value of 
$\sim$750\,\Ms\,in 
Pop~III stars per minihalo (solid lines in Fig.~\ref{fig:nlya}). This corresponds to an average SFE of 0.1\% for the streaming velocity averaged halo mass of $4.4\times10^6$\,M$_\odot$. 

\section{Conclusions}
\label{sec:conclusions}
We have shown with an idealized, semi-analytic model that Pop~III stars are crucial for establishing a
strong Ly$\alpha$-flux early in cosmic history,
which in turn can couple the spin temperature to the gas.
While we include a detailed treatment of streaming velocities, we make a number of simplifying assumptions. Our study does not address the absorption depth, in the context of interacting dark matter \citep{bar18}. Such interacting dark matter could prevent halo formation at high redshift, or heat the gas in 
halos with streaming motion between DM and baryons \citep[e.g.,][]{hb18}.

In our analysis of the Pop~III stellar component, we do not include Lyman-Werner radiation, which can delay the formation of the first stars, thus having similar consequences as streaming velocities \citep[e.g.,][]{met01, ss12}. However, an exploratory analysis has shown that streaming velocities impose the tighter constraints on 99\% of all minihalos in our fiducial model.
The interplay of Lyman-Werner
radiation and streaming velocities is not yet known, and we plan to update our analysis once quantitative estimates are available. 
Our calculation assumes a top-heavy initial mass function (IMF) for Pop~III, resulting in a higher effective temperature and thus a higher photon flux than the Pop~II counterpart. However, the lifetime of Pop~III stars is a factor of $\sim$\,3 smaller than for massive Pop~II stars, thus leading to a three-times smaller aggregate production of Ly$\alpha$-photons per stellar baryon. 
However, even with this factor of three, the Pop~II SFE would still exceed $>30$\%. 
A top-heavy IMF is thus not necessarily required to explain the EDGES signal, but a high-$z$ contribution from minihalos is. Streaming velocities suppress star formation in low mass halos, 
and therefore increase the required average stellar mass per minihalo by a factor of $\sim{}$\,5. Furthermore, our results disfavour dark matter models, which aggressively suppress the formation of small scale structures, such as axion-like ultralight dark matter \citep{sull18}, or some warm dark matter scenarios \citep{dayal17,ssb18}. 21\,cm cosmology clearly has tremendous potential to enhance our understanding of how primordial stars transformed the early Universe.
\section*{Acknowledgments}
We would like to thank Aaron Smith and the anonymous referee for their helpful comments.
Support for this work was provided by NASA through the Hubble Fellowship grant HST-HF2-51418.001-A, awarded  by  STScI,  which  is  operated  by AURA,  under  contract NAS5-26555.
VB was supported by the National Science Foundation (NSF) grant AST-1413501. 
The authors gratefully acknowledge the Gauss Center for Supercomputing for providing resources on SuperMUC at the Leibniz Supercomputing Center under projects pr92za and pr74nu.
\bibliographystyle{aasjournal}

\label{lastpage}

\end{document}